\documentstyle[epsf]{mn}
\newcommand{\hi}{H\,{\sc i} }
\newcommand{\apj}{Ap.J.}

\input epsf 
\title[Measuring non-axisymmetry in spiral galaxies]
      {Measuring non-axisymmetry in spiral galaxies}
\author[R.H.M.~Schoenmakers et al.]
        {R.H.M.~Schoenmakers,$^1$ 
         M.~Franx,$^1$
         and P.T.~de Zeeuw$^2$\\
         $^1$ Kapteyn Astronomical Institute,  
         P.O.~Box 800, 9700 AV, Groningen, The Netherlands\\
         $^2$  Sterrewacht Leiden,  
         P.O.~Box 9513, 2300 RA, Leiden, The Netherlands
         }

\date{Received date; accepted date}
\pubyear{1997}
\begin{document}
\maketitle
\begin{abstract}
  We present a method for measuring small deviations from axisymmetry
  of the potential of a filled gas disk. The method is based on a
  higher order harmonic expansion of the full velocity field of the
  disk. This expansion is made by first fitting a tilted--ring model
  to the velocity field of the gas disk and subsequently expanding the
  velocity field along each ring into its harmonic terms. We use
  epicycle theory to derive equations for the harmonic terms in a
  distorted potential. The phase of each component of the distortion
  can vary with radius. We show that if the potential has a distortion
  of harmonic number $m$, the velocity field as seen on the sky
  exhibits an $m-1$ and $m+1$ distortion.  As is to be expected, the
  effects of a global elongation of the halo are similar to an $m=2$
  spiral arm. The main difference is that the phase of the spiral arm
  can vary with radius. Our method allows a measurement of
  $\epsilon_{\rm pot} \sin 2\varphi_{2}$, where $\epsilon_{\rm pot}$
  is the elongation of the potential and $\varphi_{2}$ is one of the
  viewing angles.  The advantage of this method over previous
  approaches to measure the elongations of disk galaxies is that, by
  using \hi data, one can probe the potential at radii beyond the
  stellar disk, into the regime where dark matter is thought to be the
  dominant dynamical component.  The method is applied to the spiral
  galaxies NGC 2403 and NGC 3198 and the harmonic terms are measured
  up to ninth order.

  The residual velocity field of NGC~2403 shows some spiral-like
  structures. The harmonic analysis indicates that the $m=3$ term is
  dominant, with an average value of $\sim 0.02 v_c$. This is
  consistent with an average ellipticity of the potential of
  $\epsilon_{\rm pot} \sin 2\varphi_{2} = 0.064\pm 0.003$, but spiral
  arms may couple significantly to this result.

  In the harmonic analysis of the kinematics of NGC~3198 the $m=2$ and
  $m=3$ terms are strongest ($\sim 0.01 v_c$). The inferred average
  elongation of the potential is $0.019\pm 0.003$. Since the amplitude
  of the elongation is coupled to the viewing angles and may be
  influenced by spiral arms, more galaxies should be examined to
  separate these effects from true elongation in a statistical way.

\end{abstract}
   
\begin{keywords}
  Galaxies: individual: NGC 2403, NGC 3198 -- galaxies: kinematics and
  dynamics -- galaxies: spiral -- dark matter
\end{keywords}

\section{Introduction}

It is generally accepted that disk galaxies have massive dark halos
but little is known about their shape.  Dark halos were generally
modeled as being spherical until Binney (1978) proposed that the
natural shape of dark halos is triaxial in order to explain warped
disks.  Triaxial dark matter halos occur naturally in cosmological
N-body simulations of structure formation in the universe. These
simulations also indicate that there may well be a universal density
profile for dark matter halos (Navarro, Frenk \& White 1996; Cole \&
Lacey 1996), but the exact distribution of halo shapes is as yet
uncertain (Katz \& Gunn 1991; Dubinski \& Carlberg 1991; Warren {\it
et al.} 1992; Dubinski 1994).  The main problem is numerical
resolution, but the inclusion of baryons is also important. If baryons
are included, the baryonic disk that forms in the centre of the dark
matter distribution tends to make the halo more oblate (Barnes 1994;
Dubinski 1994). So the observed distribution of the shapes of dark
matter halos can be a constraint on scenarios of galaxy formation and
subsequent galaxy evolution.  Measuring the shapes of dark halos can
be split in two parts: measurement of the ratio $c/a$, i.e.,
flattening perpendicular to the plane of the disk, and measurement of
the intermediate to major axis ratio $b/a$, i.e., elongation in the
plane of the disk.  Polar ring galaxies provide clues on the axis
ratio $c/a$, by comparing the rotation curve of the disk with the
rotation of the inclined ring. Unfortunately, the values found for
$c/a$ are not unique, since the interpretation of the data is model
dependent, but all measurements now seem to be consistent with $c/a <
1$ (Whitmore, McElroy, Schweizer 1987; Sackett \& Sparke 1990; Sackett
{\it et al.} 1994). For our own galaxy, van der Marel (1991, and
references therein) tried to determine the flattening of the galactic
halo using counts and kinematics of halo stars. He found that the axis
ratio of the galactic dark halo must satisfy $c/a > 0.34$.  Finally,
measurements of the flaring of the \hi disks of spiral galaxies may
give us a direct handle on the flattening of the dark halo of normal
spiral galaxies. Olling (1995a) developed a method to measure $c/a$
using gas flaring, and found a very high flattening of $c/a = 0.1-0.5$
for the nearly edge-on galaxy NGC 4244 \cite{Olling}.  Preliminary
results from Sicking (1997), who developed a method for measuring the
thickness of inclined gas disks, indicate $0.2 < c/a < 0.8$ for NGC
3198 with a best value of $0.5$, whereas the measurements for NGC 2403
are difficult and give no conclusive value for the flattening of the
dark halo.  From these measurements, it seems that spherical halos can
indeed be excluded, although no good value for the typical flattening
of a dark halo has been found yet.

The elongation of the dark halo ($b/a$) is an even more undetermined
parameter. A few attempts have been made to determine the typical
elongation of disk galaxies on either photometric or kinematic basis.
On the basis of a large sample (13,482) of spiral galaxies from the
APM survey, Lambas {\it et al.} (1992) find that a pure oblate model
of spiral galaxies does not fit the observations and $b/a \sim 0.9$
fits best, consistent with the findings of Binney \& de Vaucouleurs
(1981). However, Lambas {\it et al.} remark that this effect may also
be due to the presence of spiral arms and bars in the disks. On the
basis of a distance-limited sample of 766 spiral galaxies, Fasano {\it
et al} (1992) also conclude that spiral galaxies cannot be oblate,
although early type spirals seem to have more triaxial disks than
their late type spirals. Furthermore, their late type spiral group
falls into two parts: those without bars seem to be consistent with
oblate shapes, whereas those with bars are not. These two effects may
suggest that a triaxial bulge or bar might be responsible for the
observed triaxiality, although an intrinsic elongation cannot be ruled
out by these observations.

More detailed imaging of a small sample of spiral galaxies can give a
better handle on the effects of spiral arms and bars. For this
purpose, imaging of the old stellar population in the
$\mbox{K}^{\prime} (2.2\mu)$ band was done by Rix \& Zaritsky (1995).
They observed a sample of 18 spiral galaxies and found a variety of
asymmetries in this sample. They estimated the potential in the plane
of the disk to have an average ellipticity of $0.045^{+0.03}_{-0.02}$,
but remarked that this number may only be an upper limit to the true
ellipticity, since spiral structure still couples significantly to
their estimated ellipticity.

A second way to examine elongation of the potential of disk galaxies,
is to look at the kinematics of the disks. For our own galaxy, Kuijken
\& Tremaine (1994) find on the basis of solar neighbourhood kinematics
plus kinematics of a more global kind (\hi tangent point velocities,
Cepheids, etc) that our sun lies near the minor axis of the slightly
elongated ($b/a \sim 0.9$) Milky Way disk.  Franx \& de Zeeuw (1992)
used the scatter in the Tully-Fisher relation to place limits on the
elongation of spiral disks, since elongation will increase the
scatter. Based on the observed scatter alone, they find that $b/a >
0.9$, but they argue that it is unlikely that the scatter is caused
solely by elongation and therefore the value is on average likely to
be $b/a > 0.94$.

The only direct measurement of the elongation of the potential of a
disk galaxy so far comes from the S0 galaxy IC 2006 (Franx, van Gorkom
\& de Zeeuw 1994, hereafter FvGdZ), which contains a large \hi ring in
the plane of the disk. Using information on both the kinematics and
the geometry of the ring, FvGdZ were able to measure the elongation of
the potential at the place of the ring. The measured elongation is
consistent with zero.  FvGdZ used epicyclic theory to determine the
geometry and velocity field of a ring in a mildly perturbed
potential. If the ring is in an elongated potential, the velocity
variation along the ring will not be pure circular rotation, but
higher harmonic terms will be superposed on it. Making a harmonic
expansion of the velocity field of the ring and interpreting the
measured harmonics within the framework of epicycle theory, given the
ring geometry, allows the elongation of the potential at the position
of the ring to be found. In this paper, we extend the method used in
FvGdZ for a single ring to the case of a filled gas disk in order to
study individual halo elongations for a larger sample of spiral
galaxies.

With a single ring, one has the advantage that the ring can be
considered as a single orbit. Therefore, one knows for each point in
the ring to which orbit it belongs. We will lose this advantage in the
case of a filled gas disk.  Since galactic disks are not obviously
made out of separate rings, we will decompose the \hi disks of spiral
galaxies into individual rings (the so-called tilted--ring method,
Begeman 1987) and perform a harmonic expansion on each of these rings
(Binney 1978; Teuben 1991; FvGdZ).  The harmonic expansion can be done
to arbitrary order and it produces amplitudes and phases for each
separate order.  We will investigate if we can qualitatively
distinguish between the effects of spiral arms and of elongation on
the velocity field.  We will apply this method to two spiral galaxies
with extended \hi disks, NGC 2403 and NGC 3198. From the measurements
of these galaxies, the strengths and limitations of our method will
become clear. We will see that in both cases contributions of spiral
arms to the velocity field are important, but that nevertheless limits
to the halo elongation can be set.

The outline of this paper is as follows: In Section 2 we lay down the
theoretical framework for connecting the measured kinematic terms to
perturbations of the potential. We describe the resulting procedure to
fit observed velocity fields in Section 3. In Section 4 we present
some examples of the theory, to see what effects may play a role in
interpreting the measured harmonics. Next, we examine two spiral
galaxies: NGC 2403 in Section 5 and NGC 3198 in Section 6. We conclude
with a discussion on the results from these galaxies. In the Appendix,
we give a detailed derivation of the equations presented in Section
2. Furthermore, we derive what the effect of misfitting ring
parameters (inclination, position angle and centre) is on the measured
harmonics.

\section{Description of a perturbed velocity field}

\subsection{General perturbations}

If the cold \hi gas in a disk galaxy orbits in the potential of an
axisymmetric dark halo, the resulting density field and velocity field
are axisymmetric as well. This means that the velocity field will only
show pure circular rotation. Observed velocity fields of disk galaxies
generally can indeed be fitted quite accurately by circular motion.
Many different effects can cause the remaining small residuals. Here,
we will analyse the case of a small distortion in the potential $V$,
which can be written as a sum of harmonic components, each of which
may have a different pattern speed. This is not the most general
description possible, as it does not include non-stationary phenomena
such as spiral arms. But it is a good description for the motion of
gas in a triaxial halo, since the potential distortion, i.e., the
elongation of the halo, can be considered stationary.  Furthermore, we
will assume that the gas moves in a symmetry plane on stable closed
orbits. This assumption cannot be correct in detail, since this
description of the gas motion does not allow for shocks in the gas and
therefore cannot give an accurate description of spiral arms. On the
other hand, this is a correct assumption if we are interested in the
motion of gas in a nearly circular potential \cite{tsc}, like the
potential of a mildly elongated halo. Although these assumptions
restrict the validity of our description, it will help us to gain some
understanding of the qualitative characteristics of velocity fields.

\begin{figure}
\label{geometryfig}
\epsfxsize=8.5cm
\epsfbox[5 470 580 850]{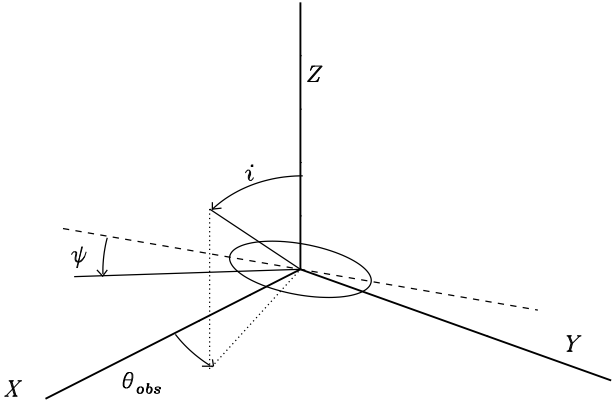}
\caption[]{Geometry of the projected orbit. The line $(\theta_{\rm
    obs},i)$ is the line-of sight. The inclination $i$ is measured
    from the $z$-axis, the axis perpendicular to the plane of the
    orbit. The angle $\theta_{\rm obs}$ is measured from the line
    $\theta=0$ and the azimuthal angle $\psi$ is measured in the plane
    of the orbit and is zero at the line of nodes, which is indicated
    by the dashed line. }
\end{figure}

The assumptions that the potential contains a small stationary
distortion and that the gas moves on stable closed orbits allow us to
use epicycle theory to analyse the velocity fields that originate from
such a perturbed potential.  We will solve the equations of motion in
this perturbed potential for the possible closed orbits and calculate
the velocity field of these orbits as it would be seen by an external
observer. In this way we can directly relate the observed velocity
field to the potential of disk galaxies.

Let $R$ and $\theta$ be polar coordinates in the rest frame of the
galaxy under consideration and consider a rotating, non-axisymmetric
potential in that frame. The potential can be written as:
\begin{equation}
\label{potentiaal1}
V(R,\theta,t) = V_0(R) + \sum_m V_m(R)\cos\{m[\theta -\Omega_{p,m}t-
\phi_m(R)]\}.
\end{equation}
Here, $V_0(R)$ represents the unperturbed potential and
$V_m(R)\cos\{m[\theta -\Omega_{p,m}t- \phi_m(R)]\}$ is the m-th
harmonic component of the perturbation, assumed to be small compared
to $V_0(R)$. The phase of the perturbation as a function of radius is
denoted as $\phi_m(R)$. Each of the harmonic components is allowed to
have its own pattern speed, $\Omega_{p,m}$. If the perturbation has
non-zero pattern speed, or $\phi_m(R)$ is not constant as a function
of radius, then the choice of the line $\theta =0$ is arbitrary.  We
focus on the velocity field as generated by a potential perturbed by a
single harmonic component. The general case, equation
(\ref{potentiaal1}), can be obtained by adding the results for
individual harmonic components, since coupling terms between different
harmonic components are of second order. For a single harmonic
component, the potential in a frame that corotates with the potential
perturbation can be written as
\begin{equation}
\label{potentiaal2}
V(R,\phi) = V_0(R) + V_m(R)\cos\{m[\phi - \phi_m(R)]\},
\end{equation}
where $\phi = \theta - \Omega_{p,m} t$.  The procedure we follow to
find the closed orbits in this potential is analogous to the procedure
followed by Binney \& Tremaine (1987, eq.\ 3-107 and further), where
they derive orbits in a potential with arbitrary $m$, but without the
radially dependent phase we introduce here. We solve the equations of
motion for a potential perturbed by a single harmonic component in
Appendix \ref{genap} and find the set of possible closed loop orbits
for any value of $m$.  Let us denote the solutions of the equations of
motion in the unperturbed potential $V_0(R)$ that represent circular
orbits by $R_0$ and $\phi_0$. If we then perturb the potential with a
single harmonic component as indicated above, we find the set of
closed loop orbits, equation (\ref{fieq}):
\begin{eqnarray}
R(\phi,t) \!\!\!&=&\!\!\! R_0 \left( 1- \textstyle{1\over 2} a_{1m}
  \cos\eta_m - a_{2m} \sin\eta_m \right), \nonumber \\ 
\phi(R,t) \!\!\!&=&\!\!\! \phi_0(t) + \textstyle{1\over 2m}(a_{1m} +
a_{3m}) \sin\eta_m - a_{4m} \cos\eta_m,
\end{eqnarray}
where $\eta_m = \eta_m(R,t) = m[\phi(t)-\phi_m(R)] \approx m[\phi_0(t)
- \phi_m(R)] = m\{[\Omega_0(R)-\Omega_{p,m}] t - \phi_m(R)\} $ and
$\Omega_0(R)$ is the usual circular frequency. The amplitudes $a_{xm}$
are defined in equations (\ref{epsR}) of the Appendix.

The resulting velocity field in the non-rotating frame for a general
rotation curve is 
\begin{eqnarray}
v_R \!\!&=&\!\! m v_c(R) [1-\omega(R)] \left( \textstyle{1\over 2}
  a_{1m} \sin\eta_m - a_{2m}\cos\eta_m \right), \nonumber \\
\label {endvelP}
v_{\phi} \!\!&=&\!\! v_c(R) \left\{ 1+\textstyle{1 \over 2}[ (1-\omega)a_{3m}+(\alpha-\omega) a_{1m}]\cos\eta_m \right. \\ \nonumber
&&\left. \phantom{\textstyle{1\over 2}} + [m(1-\omega)a_{4m} - (1- \alpha) a_{2m}] \sin\eta_m
 \right\},
\end{eqnarray}
where $\omega(R) \equiv {\Omega_{p,m} / \Omega_0(R)}$ and $v_c(R) = R
\Omega_0(R) $, the circular velocity. We have defined $\alpha = {d
\ln[v_c(R)]} / {d \ln R}$.  This velocity field is observed from a
direction $(\theta_{\rm obs},i)$, where $i$ is the inclination of the
disk and $\theta_{\rm obs}$ is defined as the angle between the line
$\theta=0$ and the observer, see Figure 1. 

Introduce $\psi = \theta-\theta_{\rm obs} + \pi/2 = \phi - \phi_{\rm
obs} + \pi/2$, where $\phi_{\rm obs}$ is the angle in the rotating
frame that corresponds to $\theta_{\rm obs}$. The angle $\psi$ is
measured along the orbit and is zero on the line of nodes. It is
sometimes called the azimuthal angle. 
It is defined fully by the 
orientation of the galaxy on the sky and  it is independent of the
internal coordinate system. 
It is straightforward to derive that the 
velocity field of a disk in pure
circular rotation is given by $v_{\rm los} = v_c(R) \sin i\cos\psi$.

Let us express the line-of-sight velocity as $v_{\rm los} = \sum_n^{}
c_n \cos n\psi + s_n \sin n\psi$. In Appendix A1 we show that the
line-of-sight velocity field has the following form:
\begin{eqnarray}
\label{lineofsight}
v_{\rm los} \!\!&=&\!\! \left[c_1 \cos \psi + s_{m-1}\sin(m\!-\!1)\psi +
  c_{m-1}\cos(m\!-\!1)\psi \right. \nonumber \\ && \left. +
  s_{m+1}\sin(m\!+\!1)\psi + c_{m+1}\cos(m\!+\!1)\psi\right],
\end{eqnarray}
where
\begin{eqnarray}
\label{termsvlos}
c_1 &=& v_* \nonumber \\
s_{m-1} &=& v_* ( -\textstyle{\frac{1}{4}}\{
[m\!-\!(m\!+\!1)\omega_m + \alpha] a_{1m} \nonumber \\
&& +(1\!-\!\omega_m)a_{3m}\}\sin m\varphi_m
\textstyle{+\frac{1}{2}}\left\{m(1\!-\!\omega_m)a_{4m} \right.\nonumber \\
&&\left. +[m(1\!-\!\omega_m)-1+\alpha]a_{2m}\right\} \cos m\varphi_m),\nonumber \\
c_{m-1} &=& v_*(\textstyle{\frac{1}{4}}\{ [m\!-\!(m\!+\!1)\omega_m+\alpha] a_{1m} \nonumber \\
&& +(1\!-\!\omega_m)a_{3m} \} \cos m\varphi_m
\textstyle{+\frac{1}{2}} \left\{m(1\!-\!\omega_m)a_{4m}
\right. \nonumber \\
&&\left. +[m(1\!-\!\omega_m)-1+\alpha]a_{2m}\right\} \sin m\varphi_m),\nonumber\\ 
s_{m+1} &=& v_* (\textstyle{\frac{1}{4}}\{
[m\!-\!(m\!-\!1)\omega_m-\alpha] a_{1m} \\
&&-\!(1\!-\!\omega_m)a_{3m}\}\sin m\varphi_m \textstyle{+\frac{1}{2}}
\left\{m(1\!-\!\omega_m)a_{4m}\right. \nonumber \\
&&\left. -[m(1\!-\!\omega_m)+1-\alpha]a_{2m}\right\}\cos m\varphi_m),\nonumber\\
c_{m+1} &=& v_* (-\textstyle{\frac{1}{4}}\{
[m\!-\!(m\!-\!1)\omega_m-\alpha] a_{1m} \nonumber \\
&&-(1\!-\!\omega_m)a_{3m}\}\cos m\varphi_m
\textstyle{+\frac{1}{2}} \left\{m(1\!-\!\omega)a_{4m} \right. \nonumber \\
&&\left. -[m(1\!-\!\omega)+1-\alpha]a_{2m}\right\} \sin m\varphi_m),\nonumber  
\end{eqnarray}
and we have written $ v_* = v_c \sin i$ and $ \varphi_m = \phi_{\rm
obs} - {\pi}/{2} -\phi_m(R)$. Now we have expressed all observable
parameters in terms of internal parameters for a potential perturbed
by a single harmonic term.

From equation (\ref{lineofsight}) we can conclude that {\it if the
potential has a perturbation of harmonic number $m$, the line-of-sight
velocity field contains an $m-1$ and $m+1$ harmonic
term}. Qualitatively, this conclusion was also inferred by Canzian
(1993).  Furthermore, we can see from the definitions of the $a_{xm}$
in the Appendix, that the amplitudes of the $m-1$ and $m+1$ terms
behave in the following way: if $\Omega_0 > \Omega_b$, the $m-1$ term
is larger in amplitude than the $m+1$ term. If $\Omega_0 < \Omega_b$,
the situation is reversed.

In general, we do not a priori know the inclination $i$ (or $q=\cos
i$), position angle $\Gamma$ and centre $(x,y)$ of the galaxy. Here,
the position angle $\Gamma$ is defined as the angle taken in
anti-clockwise direction between the north direction on the sky and
the direction of the major axis of the receding half of the galaxy.
Usually, it is assumed that the velocity field is circular, and
the values of the inclination and position angle are derived
from the best fit.
If the true velocity field contains 
non-circular terms this procedure will produce  wrong values of $i$ and
$\Gamma$ and of the position of the centre. 
If the velocity field is
expanded on the basis of these incorrect viewing angles and centre,
the harmonic terms will change. In Appendix A2 we derive the effect of
errors in these parameters: $\delta q=\delta \cos i$, $\delta\Gamma$
and centre $\delta x,\delta y$.  Under these incorrect parameters the
line-of-sight velocity becomes:
\begin{eqnarray} 
\label{incor}
v_{\rm los} \!\!\!&=&\!\!\! v_*(\hat R) \left\{(1+\alpha){\delta x\over 2R} - (1-\alpha){\delta q\over 4q}\cos\hat\psi \right. \nonumber\\
&&\!\!\!\!\!\!\!\!\!\!\left. - {\delta \Gamma \over 4}
\left[(3q+{1\over q})-\alpha ({1\over q}\!-\!q)\right] \sin \hat\psi 
  \!\!-\!\!(1\!-\!\alpha){\delta x\over 2R} 
\cos2\hat\psi  \right. \nonumber\\
&&\!\!\!\!\!\!\!\!\!\!\left.
  -(1-\alpha){\delta y\over 2qR} \sin 2\hat\psi+(1-\alpha){\delta q
  \over 4q} \cos 3\hat\psi  \right. \\
&&\!\!\!\!\!\!\!\!\!\!\left. - (1-\alpha){\delta \Gamma \over 4} ({1\over
  q}-q)\sin 3\hat\psi\right\} + \nonumber \\
&&\!\!\!\!\!\!\!\!\!\! + \sum_{i \ge 0} c_{i}(\hat R)\cos
i\hat\psi +s_i(\hat R) \sin
  i\hat\psi .\nonumber
\end{eqnarray}
where $\hat\psi$ the angle in the plane of the ring that is zero on
the apparent major axis of the ring and the $s_i,c_i$ etc. are defined
as in equation (\ref{termsvlos}).

The harmonics arising from an $m=2$ term will mix with the terms
caused by a misfit of the viewing angles. Since the tilted-ring
algorithm has no way to disentangle the effects of a misfit of the
viewing angles from that of a physical $m=2$ term, there will be first
order differences between the best fitting and true ring parameters
for $m=2$, see Appendix (A3.2). The same is true for an $m=1$ term and
the centre, Appendix (A3.1). Also, an $m=4$ term will cause $c_3,s_3$
terms, as shown in equation (\ref{lineofsight}). Due to these terms,
the position angle and inclination will change as well according to
equation (\ref{incor}), creating new $\hat c_1, \hat s_1, \hat c_3,
\hat s_3$ terms.  This affects the measurement of the elongation of the
potential and is an extra source of noise on our measurement.  In the
same way, an $m=3$ term can cause similar effects as an $m=1$ term.
For $m > 4$, the harmonics in equation (\ref{lineofsight}) do not
change to first order due to small errors in the inferred viewing angles and
centre.

\subsection{Global elongation}

We now turn to the case of a globally elongated potential, as caused
by for instance a triaxial halo. In this case we have $m=2$, $\omega
\approx 0$ and $\phi_2(R) \sim \mbox{const}$ (see also Appendix
\ref{m2}). We will assume a flat rotation curve ($\alpha=0$), since
rotation curves in the outer parts of galaxies where the dark matter
halo dominates are basically flat. Furthermore, as rotation curve
shapes approach solid body rotation ($\alpha=1$), the errors in the
tilted ring fits become large (see Appendix 3.3) and a reliable
determination of the elongation is no longer possible. In the flat
rotation curve case, the relation between the measured $a_{12}$ and
the ellipticity of the potential $\epsilon_{\rm pot}$ is $a_{12} =
\epsilon_{\rm pot}$ and $a_{32}=2\epsilon_{pot}$ (FvGdZ).

The measured line-of-sight velocity along each ring in the velocity
field will have the following form, as derived by FvGdZ (note that
their equations for $c_1$ and $s_1$ contain typesetting errors):
\begin{equation}
\label{firstvlos}
v_{\rm los} = \hat c_1 \cos\hat\psi + \hat s_1 \sin\hat\psi+\hat s_3
\sin 3\hat\psi ,
\end{equation}
with
\begin{eqnarray}
\label{fgzterms}
  \hat c_1 &=& v_*[1+\epsilon_{\rm pot} \cos 2\varphi_2], \nonumber \\ 
  \hat s_1 &=& -v_* 
    {(1-q^2)^2\over{2(1+2q^2+5q^4)}} 
      \epsilon_{\rm pot} \sin 2\varphi_2, \nonumber \\ 
  \hat c_3 &=& 0, \\
  \hat s_3 &=&
  v_* {(1-q^2)(1+3q^2)\over{2(1+2q^2+5q^4)}}
  \epsilon_{\rm pot} \sin 2\varphi_2, \nonumber
\end{eqnarray}
From the $\hat c_1$, $\hat s_1$ and $\hat s_3$ terms we can derive 
$\epsilon_{\rm
pot} \sin 2\varphi_2$. This is the quantity that we will be looking
for in our measurements of spiral galaxies.

Unfortunately, we cannot measure $\varphi_2$ separately (the $\cos
2\varphi_2$ term cannot be measured, since we do not know the exact value of
$v_*$), so all we can measure is the combination of viewing angle and
ellipticity. Global ellipticity will thus result in $s_1$ and $s_3$ terms
that are {\it constant} with radius (and thus $\epsilon_{\rm pot}
\sin 2\varphi_2$ will be constant as a function of radius as well),
whereas spiral arms will cause $s_1$ and $s_3$ terms that change sign
as a function of radius and $\epsilon_{\rm pot} \sin2\varphi_2$
will wiggle, as we shall see in Section \ref{examfields}.  Radial
variation in $\epsilon_{\rm pot} \sin 2\varphi_2$ may help us to
distinguish between the effects of spiral arms and systematic
elongation.

Equation (\ref{fgzterms}) shows that the $c_3$ term should always be
zero. If the inclination is not equal to the best fitting inclination,
the $c_3$ terms will not be equal to zero. Therefore, monitoring $c_3$
in the fitting procedure will be a useful way to check the correctness
of the fitted inclination.  Furthermore, spiral arms will also
influence the measurements of the inclination. As shown in the
Appendix, if $\varphi_2(R)$ is constant, the fitted inclination for
each ring is identical to the true inclination for that ring. But, if
$\varphi_2(R)$ varies as a function of radius, the fitted inclination
will deviate from the true inclination. This effect will not be large
(typically one or two degrees), but it will be measurable.

\section{Fitting velocity fields}
\label{fitvel}
\subsection{The method}

In order to determine the variation of different harmonic components
as a function of radius, we want to fit the velocity field of a galaxy
along individual, independent rings. Within the {\sc GIPSY} package
(Groningen Image Processing SYstem, van der Hulst {\it et al.}, 1992)
there is a routine that fits a set of so-called tilted--rings to the
velocity field of a galaxy. The code fits a circular model to the
velocity field by adjusting the ring parameters (kinematic centre
$(x_0,y_0)$, inclination $i$, position angle $\Gamma$ and systemic
velocity $v_{\rm sys}$), so we have a list of ring parameters as a
function of radius.  All the ring parameters are simultaneously fitted
with a general least-squares fitting routine.  In the tilted--ring
fit, we should take a uniform weighting function since points near the
minor axis contain information about ellipticity. It is best to keep
the centre fixed in the fit because the galaxy may contain physical
$c_2,s_2$ harmonics in its velocity field. In that case, a free centre
would drift in such a way as to make these terms disappear.  If a
galaxy would be truly lop-sided in its kinematics (i.e., the kinematic
centre drifts as a function of radius), this would show itself in the
$c_2$ and $s_2$ terms, as can be seen from equation (\ref{obserm1}).

After the tilted ring decomposition we make a harmonic expansion of
the radial velocity along each ring (using the ring's individual
$\Gamma, i, x_0,y_0$):
\begin{equation}
\label{expansion}
v_{\rm los} = v_{\rm sys} + \sum_{n=1}^{k} \hat c_n \cos n \hat\psi +\hat s_n
\sin n \hat\psi,
\end{equation}
where $k$ is the order of the fit.  The coefficients $\hat c_n, \hat
s_n $ are determined by making a least-squares fit with a basis
\begin{equation}
\label{basisfunction}
\left\{1,\cos \hat\psi,\sin\hat\psi,\cdots,\cos k \hat\psi, \sin k
  \hat\psi \right\}, 
\end{equation}
\noindent
on the data points in each ring.

We assume here that we can identify each pixel in the velocity map
with a unique position in the galaxy. Strong warps in the outer
\hi-layers for instance tend to invalidate this assumption, since the
value of the velocity field in a certain pixel may be composed of two
or more rings (the line-of-sight crosses 2 or more rings) and the
velocity field will not be accurate there.  Mild warping where the
viewing angles change by only a few degrees will in general not be a
problem, as long as individual rings do not overlap each other.

\subsection{Error estimates}

The errors given by the tilted--ring decomposition code and by the
harmonic fitting procedure are formal errors of the fits and do not
take correlation between points due to beam smearing into
account. Sicking (1997) has derived that the formal errors should be
multiplied by a factor
\begin{equation}
\label{beamfloor}
\beta = \sqrt{4\pi \sigma_x \sigma_y \over{\delta x \delta y}},
\end{equation}
where $\delta x,\delta y$ are the dimensions of a pixel in arcsec and
$\sigma_x,\sigma_y$ are the dispersions of the beam in arcsec. All
measurement errors presented in this paper are corrected for this
correlation effect.

\section{Some example velocity fields}
\label{examfields}

\begin{figure}
\label{figEXAMPLE}
\epsfxsize=8.5cm
\epsfbox[100 375 470 660]{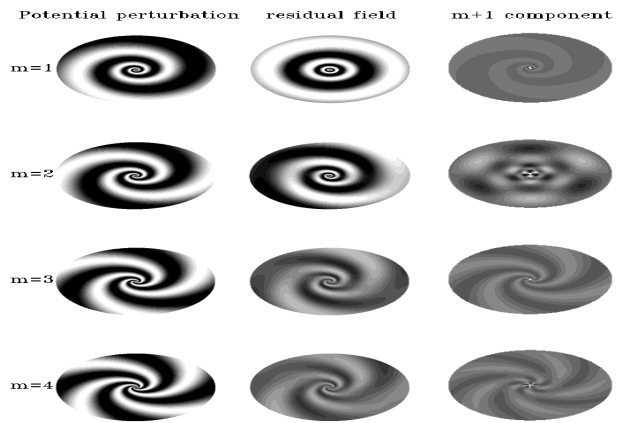}
\caption[]{Potential perturbations and decomposed velocity fields. From top to
  bottom: the $m=1, 2, 3$ and 4 terms, from left to right : a. The potential
  perturbation, b. the residual velocity field (total velocity field
  minus circular velocity), showing mainly the $m-1$ component. No
  tilted-ring fit had to be made to find the circular velocity, since
  the circular velocity is a known input paramenter in these models.
  c. The residual field minus the $m-1$ component, revealing the $m+1$
  component. To find this component a harmonic fit has been made. 

  Normally, when creating residual fields, one would also subtract the
  kinematic $m=0$ term. In that case, the kinematic $m=0$ term in the
  top row, middle column, would be zero and one would only see the
  kinematic $m=2$ term. If the centre would be a free parameter in the
  tilted--ring fit, any kinematic $m=2$ component in the residual
  field would be zero as well. Finally, the alternating behaviour of
  the $m=3$ kinematic term (second row, third column) is due to the
  fact that in the tilted--ring fit the inclination is a free
  parameter. Therefore, the $c_3$ term is fitted to zero and only the
  $s_3$ term is visible.\par\noindent All plots are on the same scale.
  The potential ranges from $-80$~km$^2$~s$^{-2}$ to
  $80$~km$^2$~s$^{-2}$ , all velocity components from $-2$~km~s$^{-1}$
  to $2$~km~s$^{-1}$. White denotes positive, black negative values.}
\end{figure}

In order to clarify the equations presented in Section 2 and in
Appendix A, we present some example velocity fields in Figure 2. The
different terms range from $m=1$ to $m=4$.  For the phase of the
fields, $\varphi_m(R)$, we chose a logarithmic spiral, $\varphi_m(R)=
\ln({R}/{R_{\rm s}})/ \tan p$. The amplitude of the perturbation has been
taken constant throughout the field.  In Table \ref{tableparam}, we
summarise the parameters used to create the model fields. The fields
are then created using equations (\ref{lineofsight}) and
(\ref{termsvlos}).  Subsequently, these line-of-sight velocity fields
were fitted as described in Section 3, revealing the two harmonic
components that were hidden in them.

The example velocity fields are shown in Figure 2. We display from
left to right
\begin{enumerate}
\item{The potential containing the $m$ term perturbation.}
\item{The residual velocity field caused by this potential perturbation
      (i.e., the $m-1$ plus $m+1$ terms together).}
\item{The $m+1$ component of the velocity field.}
\end{enumerate}
\begin{table}
  \begin{center}
    \leavevmode
    \caption{Parameters used for model fields}
    \label{tableparam}
    \begin{tabular}{|l|l|l|}
quantity & symbol & value \\ \hline
inclination & i & 60\degr \\ 
obs. angle & $\phi_{\rm obs}$ & $45\degr$ \\ 
pattern speed & $\Omega_{p,m}$ & 0 km~s$^{-1}$~kpc$^{-1}$ \\
amplitude & $V_m$ & 80 km$^2$~s$^2$ \\
circular velocity & $v_c$ & 150 km~s$^{-1}$ \\ 
maximum radius & $R_{max}$ & 18 kpc \\ 
spiral scale length & $R_s$ & 10 kpc \\
pitch angle & p & $10\degr$ \\ \hline
    \end{tabular}
    
  \end{center}
\end{table}

We see that indeed an $m$ term in the density causes $m-1$ and $m+1$
terms in the kinematics of the gas. Since the pattern speed $\Omega_p$
was taken zero for all models, the $m-1$ term will be dominant, see
Section 2.  The $m=2$ potential perturbation
demonstrates the effect of fitting a circular model to a non-circular
velocity field. The fitting procedure adapts the inclination until
$c_3=0$ so that only the $s_3$ term remains. In our example, the $s_3$
term varies as a function of radius.  If the field would have
contained a global intrinsic ellipticity, $m=2,
\varphi_2(R)=\mbox{const}$, the $s_1$ and $s_3$ terms would have been
constant as a function of radius. Therefore, the measured $s_3$ would
have been constant and non-zero as well, but no radially alternating
pattern for $s_3$ would have been visible.
 
Since in the fitting procedure the centre was kept fixed, we also see
a two-armed spiral in the $m=1$ case. If we would have taken the
centre as a free parameter in the fit, it would have drifted in such a
way as to make the $c_2$ and $s_2$ terms disappear.

We shall encounter the effects discussed above again while
interpreting the data of NGC 2403 and NGC 3198.

\section{NGC 2403}

\subsection{Data description}

\begin{figure*}
\label{2403depro}
\epsfxsize=18cm
\epsfbox{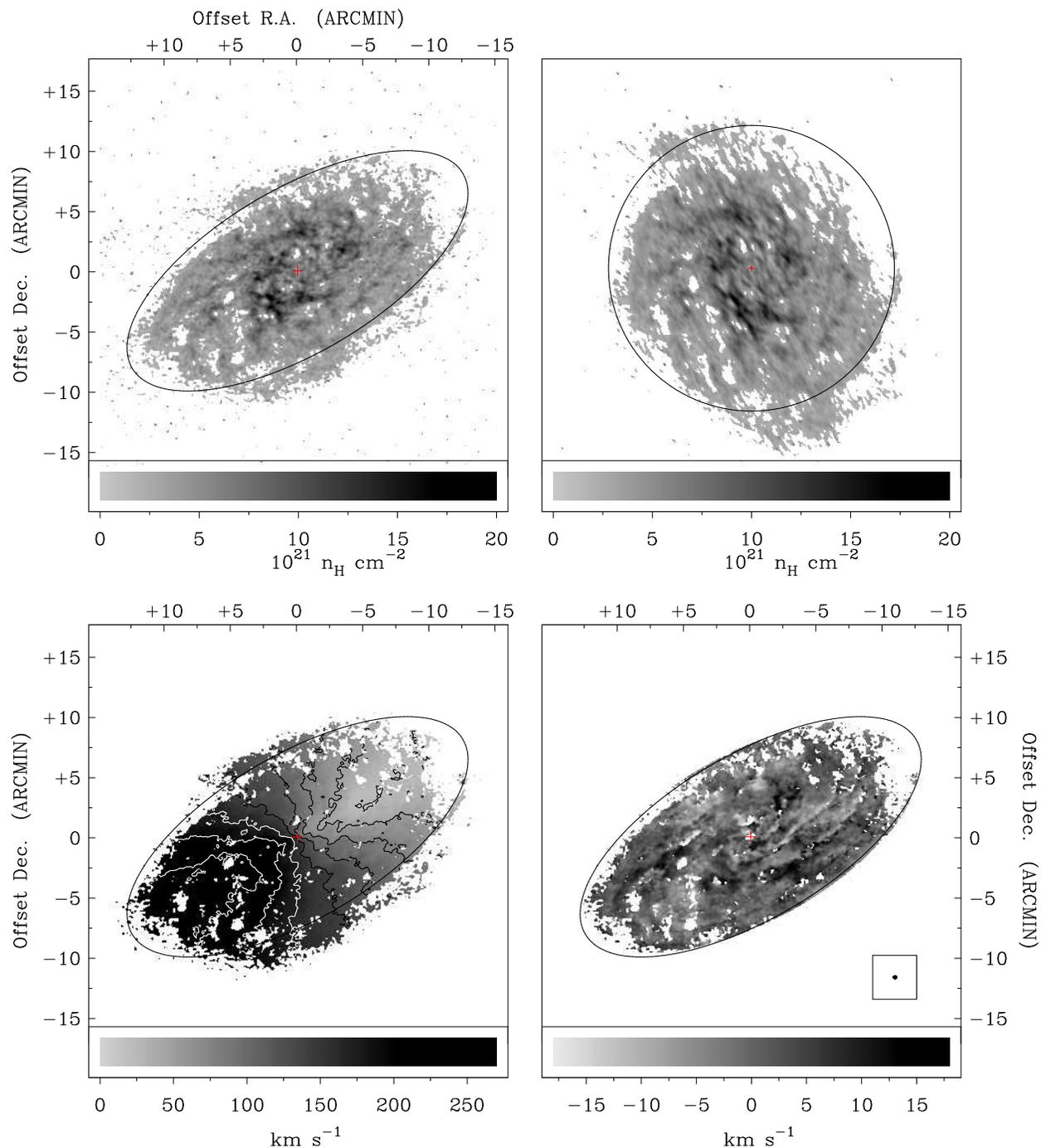}
\caption[]{WSRT observations of NGC 2403.  Top left: Map of the \hi
  surface density. Some spiral-like structure is visible. The ellipse
  is the outermost ring fitted to the data in the tilted--ring
  decomposition. Top right: The deprojected \hi image of NGC 2403
  (deprojected with constant inclination). Bottom left: The velocity
  field.  Systemic velocity is at 133.5 km~s$^{-1}$. The contours
  range from 20 km~s$^{-1}$ to 260 km~s$^{-1}$ and are spaced 30
  km~s$^{-1}$ apart. Bottom right: The
  residual velocity field.  Here we have subtracted the result of a
  tilted--ring fit from the original velocity field. Some structure
  can be seen. Note the small amplitude of the residuals (less than
  10\%) with respect to the original field. The beam size is indicated
  in the box in the lower right corner.}
\end{figure*}

NGC 2403 is a nearby (3.25 Mpc Begeman 1987) Sc(s)III galaxy
\cite{tamm}. It has been observed $4 \times 12$ hours with the WSRT
\cite{floor}. The data have been smoothed to a circular beam of 13
arcseconds. With this beam and a pixel size of $5 \times 5.5$
arcseconds, the $\beta$-factor of equation (\ref{beamfloor}) is
$3.73$.

NGC~2403 is well suited for our analysis:
\begin{enumerate}
\item{It has a large extent on the sky (Holmberg dimensions $29^{\prime}
       .0 \times 15^{\prime} .0$) }
\item {It has a favourable inclination ($i \approx 62\degr$)}
\item {The data is of high signal/noise and has a small beam}
\item {The galaxy shows no obvious warp}
\end{enumerate}

Figure 3 shows the surface density map of the \hi gas in the upper
left corner. All data with a $\mbox{S/N} < 3$ is clipped away, in
order to have a reliable velocity field.  The ellipse is the outermost
ring ($R = 858$ arcsec) fitted to the velocity and surface density
fields. Outside this radius, the data has suffered too much from
clipping to make a sensible harmonic decomposition possible. The
centre of all the rings is indicated by a red cross and was kept fixed
during the tilted--ring fit.

Next to this map we present the deprojected \hi surface density
map. We took an inclination of $62\degr$ to deproject the galaxy, the
average value found by the tilted--ring fits. NGC 2403 seems to be a
bit elongated in the direction of the (apparent) minor axis. Some 
traces of spiral arms can be seen. 

At the bottom left the velocity field is shown. It is relatively
regular apart from some small wiggles in the iso-velocity contours,
probably caused by spiral arms.  Next to the velocity map is the
residual velocity field, created by subtracting the fitted circular
velocity from the true velocity field. The residuals are not large ($<
10$ km~s$^{-1}$ in absolute magnitude) and appear to show some
systematic, three-fold, structure. The beam shape is indicated in the
lower right corner.

\subsection{Results of the harmonic expansion}

\begin{figure*}
\label{2403param}
\epsfxsize=18cm
\epsfbox{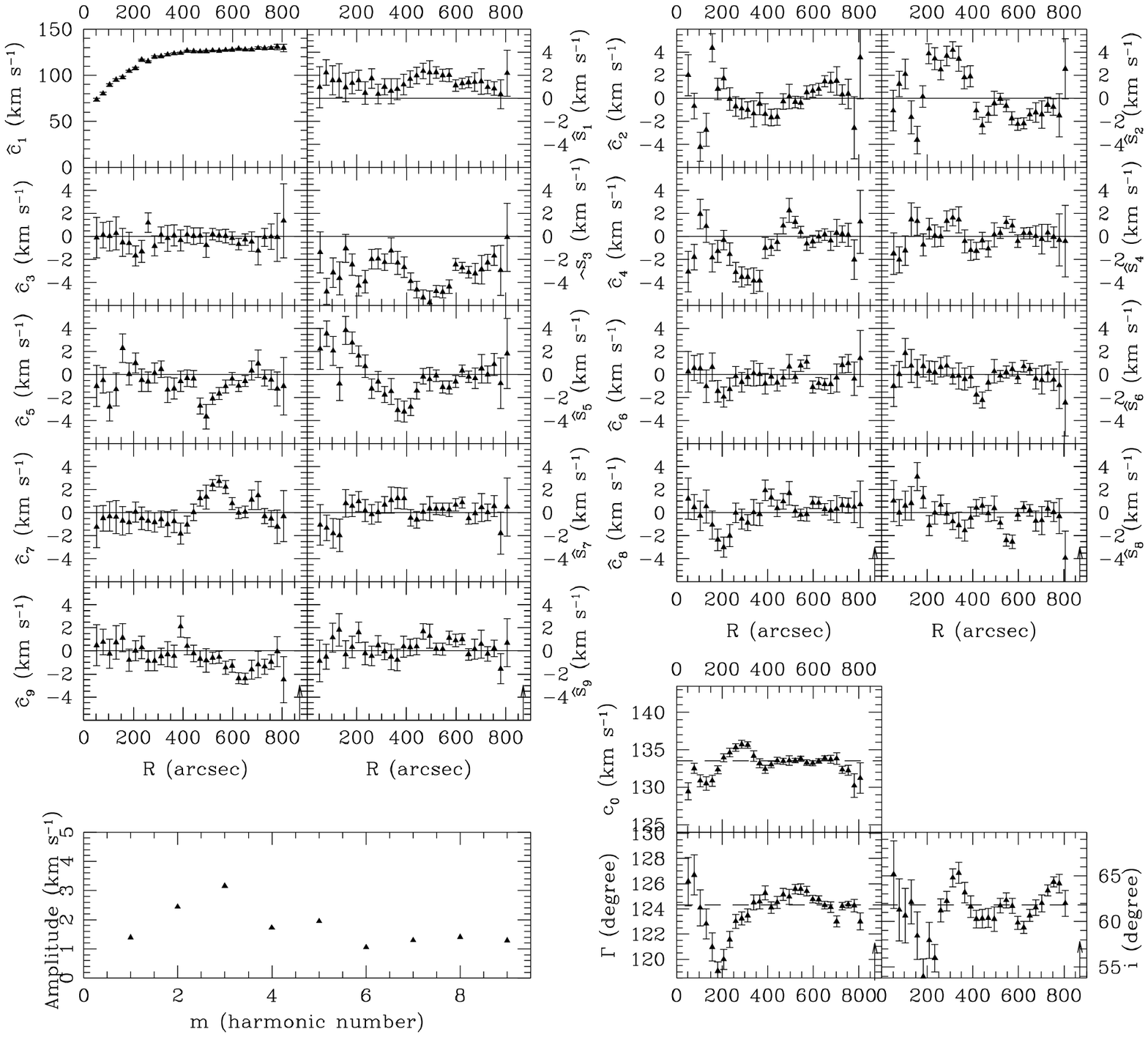}
\caption[]{Harmonic expansion and ring parameters of NGC 2403 as a
  function of radius. Average values of the ring parameters are
  indicated by a dashed line.  The points are separated by 26 arcsec
  ($2$ beams) and are therefore independent. The Holmberg radius is
  indicated by an arrow at the end of the box (at 870 arcsec).  The
  amplitude of the $\hat s_3$ term is the largest with an amplitude
  $\approx 3$ km~s$^{-1}$. This is about $2\%$ of the rotation
  velocity $\hat c_1$. At the bottom the average amplitude ($A_i =
  \sum_j (\hat c_i(j)^2 + \hat s_i(j)^2)^{1/2} / j$, $A_1 = \sum_j
  \sqrt{\hat s_1^2(j)}/j$) of the harmonic terms is plotted.  
}
\end{figure*}

The first step in our procedure is to expand the velocity field of
NGC~2403 into individual concentric tilted--rings. The parameters $i$,
$\Gamma$ and $v_{\rm sys}$ are measured locally for each ring, in
order to be able to use the exact formalism of Appendix A3. We took
the width of each ring to be 26 arcseconds, i.e., 2 beams. Therefore,
the individual points in the plots of Figure 4 can be considered
independent. Figure 4 presents the inclination, position angle and
systemic velocity of each ring as a function of radius.  The
fluctuations in the inclination $i$ and position angle $\Gamma$ are
most probably caused by an $m=2$ spiral arm in the density. In the
case of $\varphi_2(R) \neq 0$, the measured inclination will in
general not be equal to the true inclination, as mentioned
before. This will cause the measured inclination to wiggle around its
average value. We expect this average value to be very close to the
true inclination. The fluctuations in the systemic velocity, $\hat
c_0$, are of comparable magnitude as the $\hat c_2,\hat s_2$ terms, as
it should be according to Appendix 3.1 if these variations are
caused by an $m=1$ term in the surface density. It should resemble
$-\hat c_2$, but if $\hat c_2$ is also influenced by an $m=3$ term in
the density, this need not exactly be the case.  The weighted average
values of the ring parameters are:
\begin{enumerate}
\item{position angle: $124\fdg 2 \pm 0\fdg 04$}
\item{inclination: $61\fdg 8 \pm 0\fdg 18$}
\item{systemic velocity: $133.5 \pm 0.09$ km~s$^{-1}$}
\end{enumerate}

In Figure 4 we also present the result of the harmonic expansion of
the tilted--rings. The arrow just beyond the last measured point
denotes the Holmberg radius. This means that our \hi observations are
completely within the optical disk of NGC 2403.

After an initial rise, the rotation curve of NGC 2403 is relatively
flat within errors. The $\hat c_3$ term has been fitted to zero by the
tilted--ring fit, indicating that the inclination has been fitted
correctly to the rings (but note again that in the case of an $m=2$
spiral arm, this inclination is not identical to the true inclination
of the ring).  There is some $\hat c_4$ at radii smaller than $400$
arcsec that may be caused by an $m=3$ component in the potential. This
$m=3$ component may also influence the $\hat c_2$ and $\hat s_2$
harmonics.  These $\hat s_2$ and $\hat c_2$ terms are relatively large
and are probably caused by an $m=1$ component in the potential.  The
$\hat s_3$ term is the strongest harmonic found in NGC 2403. This term
is negative at all radii. We can calculate $\epsilon_{\rm pot}
\sin2\varphi_2$ from $\hat s_1$ and $\hat s_3$ at each radius as (cf.\
eq.\ [9])
\begin{equation}
\label{calceps}
\epsilon_{\rm pot} \sin 2\varphi_2 = {(\hat s_3 - \hat s_1)
  {(1+2q^2+5q^4)\over \hat c_1 (1-q^4)}}
\end{equation}

Figure 5 shows $\epsilon_{\rm pot} \sin 2\varphi_2$ as a function of
radius. It shows significant variations, but is negative at all
radii. As we saw in the numerical examples of the previous paragraph,
spiral arms would cause a radially alternating pattern of positive and
negative values for the $\hat s_3$ kinematic term, whereas a global
elongation would cause the $\hat s_3$ term to be constant and non-zero
throughout the disk. A possible explanation for the behaviour of the
measured $\hat s_1$,$\hat s_3$ is therefore that the average value of
$\epsilon_{\rm pot} \sin 2\varphi_2$ is caused by a global elongation
of the disk and that the wiggles superposed on it are due to spiral
arms. The weighted average value of $\epsilon_{\rm pot}
\sin 2\varphi_2$ is $-0.064 \pm 0.003$. Since $\epsilon_{\rm pot}
\sin 2\varphi_2$ has only physical meaning in the case of a more or
less flat rotation curve, it should only be trusted outside about 200
arcsec. Harmonics higher than $m=5$ have small amplitudes.
\begin{figure}
\label{2403graph}
\epsfxsize=8.5cm
\epsfbox{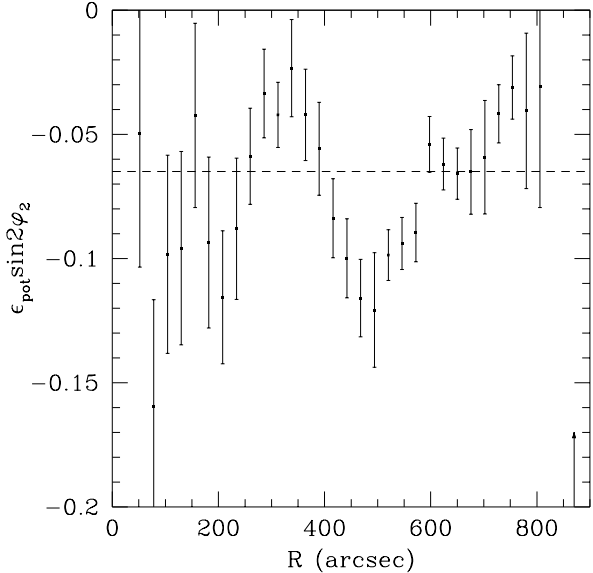}
\caption[]{The elongation measurement as calculated from the $\hat
  s_1$,$\hat s_3$ and inclination (eq.\ \ref{calceps}). It varies
  as a function of radius, but is negative throughout the galaxy. A
  possible explanation for this behaviour is that the average value of
  $\epsilon_{\rm pot} \sin 2\varphi_2$ is caused by some global
  elongation and the wiggling is caused by other effects like spiral
  arms. The average value is $-0.064 \pm 0.003$ }
\end{figure}

\section{NGC 3198}

\subsection{Data description}

\begin{figure*}
\label{3198data}
\epsfxsize=18cm
\epsfbox{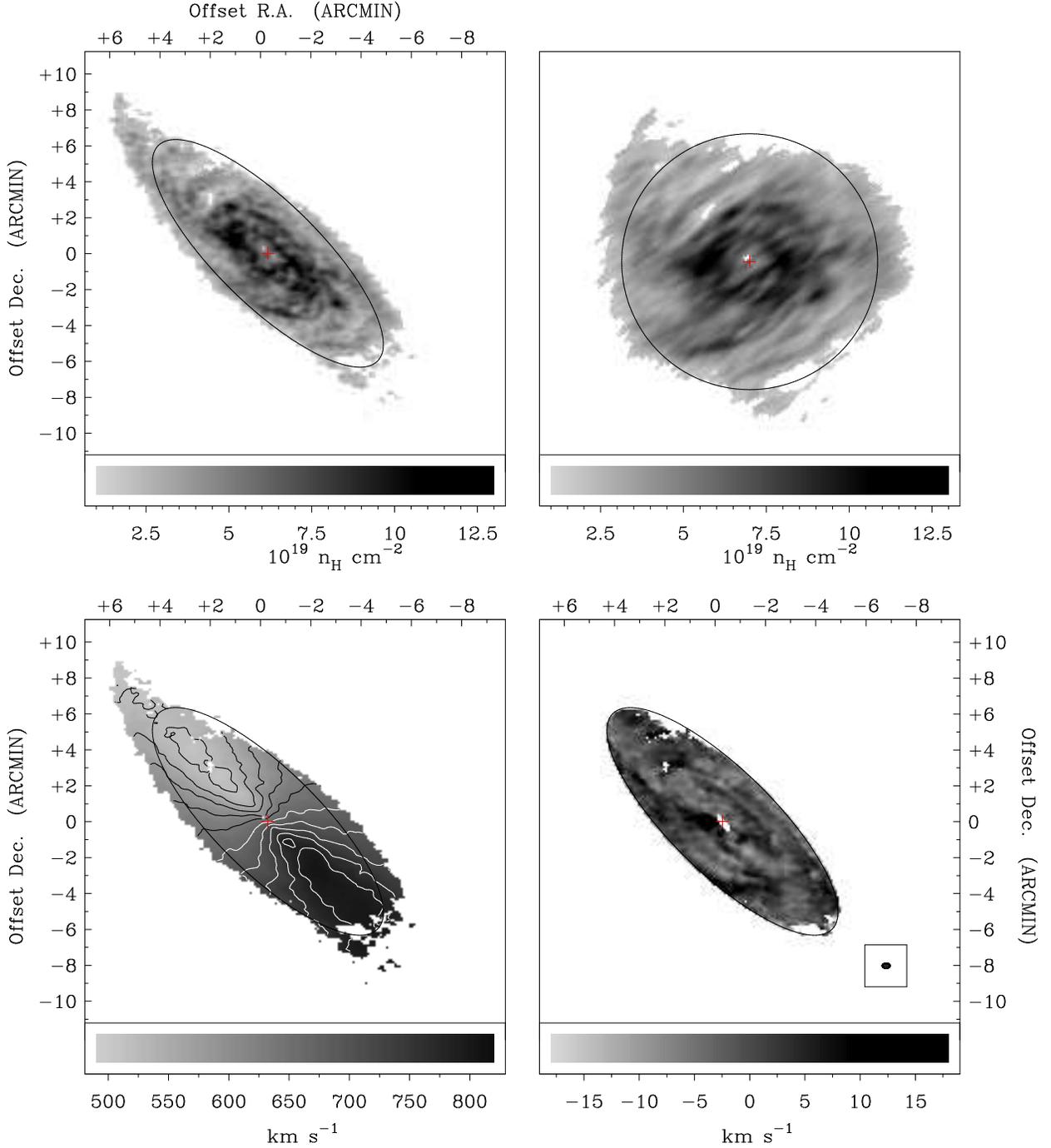}
\caption[]{WSRT observations of NGC 3198.  Top left: Map of the \hi
  surface density. Some spiral-like structure is visible outside an
  \hi ring-like structure. The ellipse is the outermost ring that has
  been fitted to the data. Top right: The deprojected \hi image of NGC
  3198 (deprojected with constant inclination). The structure and \hi
  ring is better visible here. Bottom left: The velocity field. The
  systemic velocity is at 659.6 km~s$^{-1}$. The contours
  range from 520 km~s$^{-1}$ to 790 km~s$^{-1}$ and are spaced 30
  km~s$^{-1}$ apart. Bottom right: The
  residual velocity field. Here we have subtracted the result of a
  tilted--ring fit from the original velocity field. There is some
  four-fold structure visible. The beam size is indicated in the box
  in the lower right corner.}
\end{figure*}

The second relatively nearby (9.4 Mpc, Begeman 1987) spiral galaxy we
will examine is the Sc(rs)I-II \cite{tamm} galaxy NGC~3198. This
galaxy was observed $4 \times 12$ hours with the WSRT \cite{floor}.
The observations were subsequently reprojected to a circular beam with
FWHM of $18$ arcsec. The $\beta$ factor, equation (\ref{beamfloor}),
for NGC 3198 is $4.6$.

NGC 3198 also meets our criteria:
\begin{enumerate}
\item{It has large dimensions on the sky (Holmberg dimensions $11^{\prime} .9 \times 4^{\prime} .9$)}
\item{The inclination is $i \approx 71 \degr$}
\item{The data is of high S/N and has a small beam}
\item{NGC 3198 has no warp of significance}
\end{enumerate}

In Figure 6 we present the data for this galaxy. The data is of
somewhat better quality than the data for NGC 2403, since the disk of
NGC 3198 is completely filled, although the beam is larger. The
surface density field shows a ring-like structure of a few arcminutes,
with 3 spiral-like structures extending from it. The ellipses and the
circle again indicate the outermost data used for the harmonic
analysis ($450$ arcsec in this case). The cross indicates the position
of the centre of the rings. The velocity field of NGC 3198 is quite
regular and the amplitudes of the residual field are not very large,
typically less than 10 km~s$^{-1}$ in both the positive and negative
direction. Even though these residuals are small, they do show some
systematic structure. It is worth noting that the residuals are
strongest near the centre of NGC 3198.
 
\subsection{Results of the harmonic expansion}

\begin{figure*}
\label{3198param}
\epsfxsize=18cm
\epsfbox{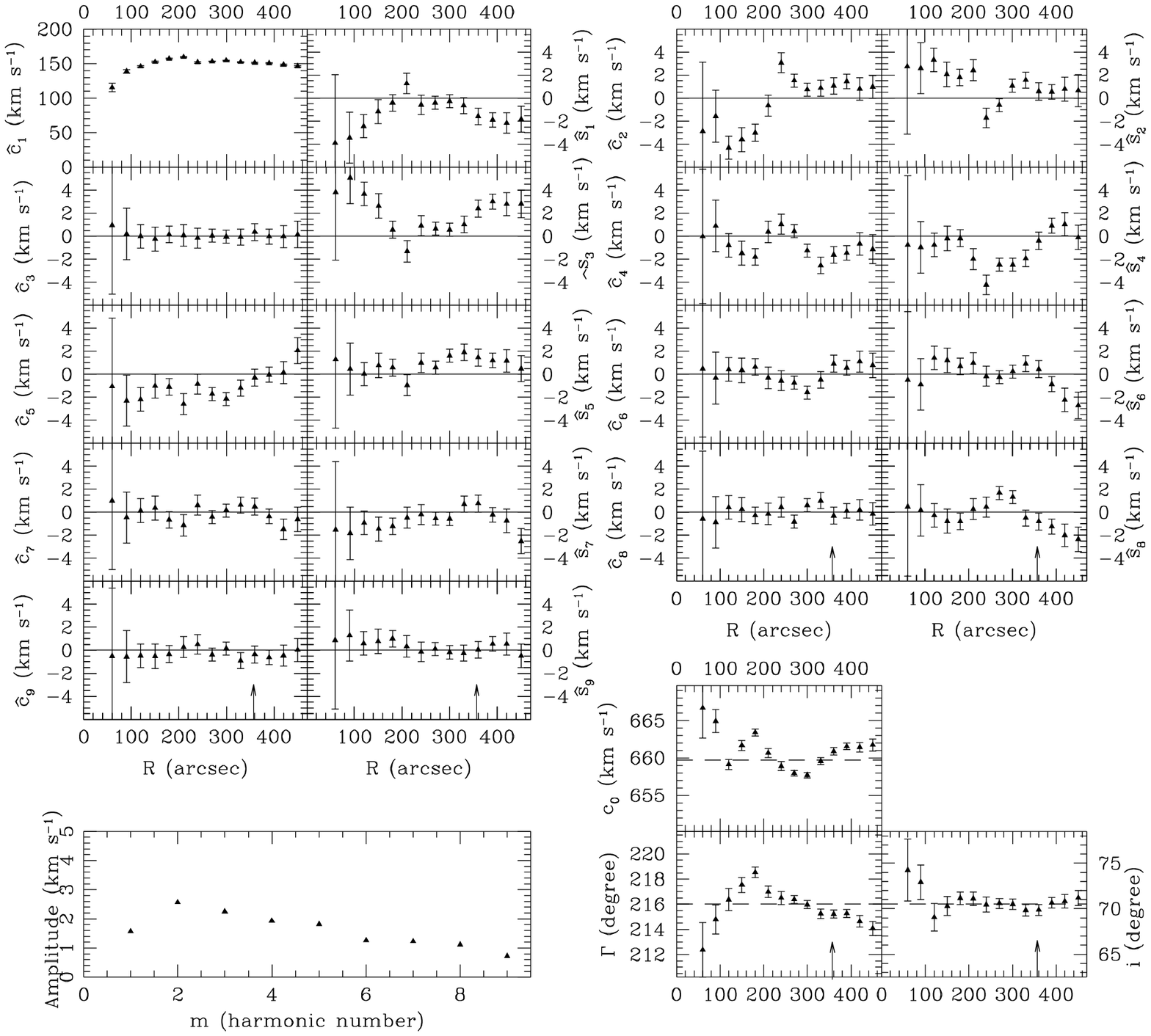}
\caption[]{Harmonic decomposition and ring parameters of NGC 3198 as a
function of radius.  Average values are indicated as a dashed
line. The points are seperated by $1.7$ beams and are therefore
independent. The Holmberg radius is indicated by an arrow. The total
amplitude, bottom graph, decreases as a function of harmonic number.
The systemic velocity $\hat c_0$ varies a few km~s$^{-1}$ over the
galaxy, comparable in amplitude with the largest harmonics. The
position angle $\Gamma$ varies a few degrees as a function of radius.
The inclination $i$ is constant within measurement errors. }

\end{figure*}

We fitted a set of concentric rings to the velocity field of NGC 3198.
We present the results for the ring parameters in Figure 7. The systemic velocity, $c_0$, varies with an amplitude of a few km~s$^{-1}$, about
the same amplitude variation as the strongest terms in the harmonic
decomposition.  The inclination of the
rings, $i$, is constant within errors. The position angle $\Gamma$ varies a few degrees. The weighted values of the ring
parameters found are:
\begin{enumerate}
\item{position angle: $216\fdg 0 \pm 0\fdg 05$}
\item{inclination: $70\fdg 5 \pm 0\fdg 20$}
\item{systemic velocity: $659.6 \pm 0.15$ km~s$^{-1}$}
\end{enumerate}
Figure 7 also presents the result of the subsequent harmonic fit.
After an initial rise, the rotation curve is flat within errors up to
the last measured point at 450 arcsec. This is not the outermost limit
of the observed \hi distribution (see Figure 6) but since the disk
shows big gaps at radii larger than 450 arcsec, the harmonic expansion
would become too unreliable. The arrows denote the Holmberg radius, so
our measurements go out to about 1.3 Holmberg radii.

As predicted by the theory, the $\hat c_3$ term has been fitted to
zero by the tilted--ring fit. Together with the fact that the
inclination is constant as a function of radius, this tells us that
NGC 3198 does not contain strong $m=2$ spiral arms, although a global
elongation of the disk is still possible. The $\hat c_2,\hat s_2$
terms are quite strong in the inner part of NGC 3198. The centre has
been fixed at such a value as to minimize the $\hat c_2,\hat s_2$
terms. As a consequence the $\hat s_2,\hat c_2$ terms are relatively
small in the outer part of NGC 3198.  If the centre is left as a free
parameter in the tilted-ring fit, it drifts by a few hundred parsec in
order to make the $\hat c_2,\hat s_2$ terms zero. In other words, NGC
3198 is somewhat lop-sided.

In the surface density map we saw that there appear to be three
spiral-like structures in the \hi. This would cause some non-zero $m=2$ and
$m=4$ terms in the kinematic expansion of the velocity field. Indeed,
we see that there is some power in the $\hat c_4,\hat s_4$ terms that
may be caused by an $m=3$ component in the \hi field.

Finally, there is a strong $\hat s_3$ term, together with a less
strong $\hat s_1$ term. From these two terms we deduce $\epsilon_{\rm
pot} \sin 2\varphi_2$ as a function of radius, as plotted in Figure 8.
The low amplitude of $\epsilon_{\rm pot} \sin 2\varphi_2$ is striking:
its weighted average is $-0.019\pm 0.003$. This low value may be
caused by $\epsilon_{\rm pot}$ itself being small, or the viewing
angle $\varphi_2$ being such as to make $\sin 2 \varphi_2 $ much
smaller than unity. The variation in $\epsilon_{\rm pot} \sin
2\varphi_2$ can be due to the variation of $\epsilon_{\rm pot}$ with
radius, or $\varphi_2(R)$ is not completely zero, or a combination of
both.

\begin{figure}
\label{3198graph}
\epsfxsize=8.5cm
\epsfbox{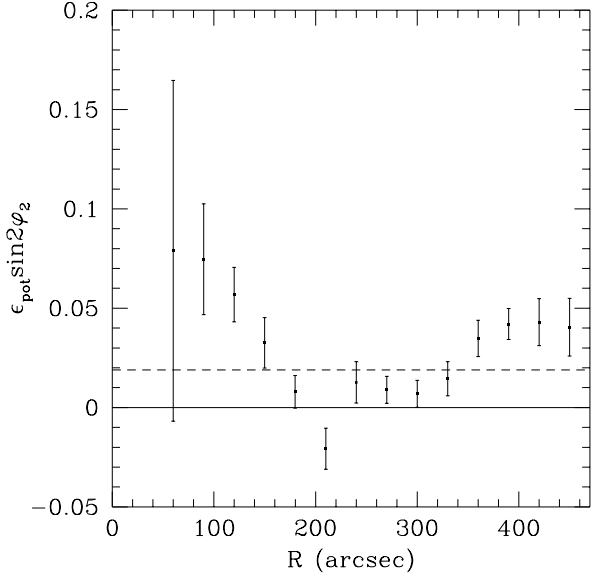}
\caption[]{Elongation $\epsilon_{\rm pot} \sin 2\varphi_2$, as
  measured from the $\hat s_1$, $\hat s_3$ terms and inclination. Its radial
variation is not very large within errors, indicating that global
ellipticity might be the cause of the non-zero measurement, and its
average amplitude is low ($-0.019 \pm 0.003 $). 
The arrow indicates the Holmberg radius.}
\end{figure}

\section{Summary and discussion}

We have extended the FvGdZ formalism for measuring elongation of
potentials to the case of a slightly non-axisymmetric filled gas disk,
which may contain spiral-like perturbations. All the details of this
derivation can be found in the Appendix. This analysis assumes a
stationary perturbation in the frame rotating with the perturbation,
and closed stable orbits.  According to this analysis, a perturbation
in the potential of harmonic number $m$ causes an $m-1$ and $m+1$
perturbation in the radial velocity field. The $m-1$ component will be
dominant in amplitude if $\Omega_0 > \Omega_b$, otherwise the $m+1$
component will be dominant.  Applicability to non-linear phenomena
like spiral arms is limited. But a small global elongation of the
overall potential, as is the case with a triaxial dark matter halo,
can be analysed with our method.  In the case of a triaxial potential,
we have $m=2$, small pattern speed and no radial dependence of the
phase of the $m=2$ perturbation. If we assume a flat rotation curve,
we can deduce $\epsilon_{\rm pot} \sin 2\varphi_2 $. Here,
$\epsilon_{\rm pot}$ is equal to the ellipticity of the potential and
$\varphi_2$ is the angle between the minor axis of the elliptical orbit
plus a phase term and the viewing angle.
 The term $\epsilon_{\rm pot}
\cos 2\varphi_2$ cannot be determined.  In order to measure this
$\epsilon_{\rm pot} \sin 2\varphi_2 $ and other perturbations of
velocity fields, we need to expand the gas disk into individual
rings. This was done by fitting a tilted--ring model to the velocity
fields. A tilted-ring model tries to fit a circular model to the
velocity field. Given the rings found by the tilted--ring fit, we made
a harmonic expansion of the velocity field along each ring. In that
way we obtain the harmonic terms $\hat c_n,\hat s_n$ as a function of
$R$, where we took $n \le 9$ for our fits. Given $\hat s_1(R)$ and
$\hat s_3(R)$, we can calculate $\epsilon_{\rm pot} \sin 2\varphi_2 $
as a function of radius. Other perturbations of the velocity field,
like spiral arms, do not affect the elongation measurement, unless
they contain some $m=2$ or $m=4$ component in the potential. In that
case, the global elongation and the other perturbing effect cannot be
uniquely disentangled. We propose that the absolute mean value of
$\epsilon_{\rm pot} \sin 2\varphi_2$ is an upper limit of the true
elongation and the wiggling is caused by spiral arms.  Since the
method relies on the fact that the velocity in each point in the
velocity field is uniquely determined, the method cannot be applied in
those parts of galaxies where strong warping is evident.

Two spiral galaxies with regular \hi velocity fields were analysed
with this method: NGC 2403 and NGC 3198. Both galaxies show less power
in the higher order ($m>5$) terms than in the lower order terms.

NGC 2403 clearly shows spiral-like behaviour. Its $\hat s_3$ term is
large and negative at all radii and the amplitude changes as a
function of radius. Two effects can contribute to this measurement: a
global elongation of the disk causing a constant non-zero $\hat s_3$
and spiral arms, as are visible in the surface density map, that cause
the wiggles. The radially averaged $|\langle \epsilon_{\rm pot} \sin
2\varphi_2 \rangle|$ is $0.064 \pm 0.003$.

NGC 3198 shows signs of spiral arms in the $m=4$ kinematic mode (in
the \hi surface density map a three-armed spiral-like structure can be
seen).  The relatively large $\hat s_2,\hat c_2$ terms in the central
parts may be caused by lop-sidedness of the galaxy and the $\hat
s_1,\hat s_3$ terms are most probably caused by global elongation. The
deduced value for $|\langle \epsilon_{\rm pot} \sin 2\varphi_2 \rangle|$ is
$0.019 \pm 0.003$.

We conclude that both galaxies may show signs of small global
elongation ($< 0.1$), although it is hard to separate the effects of
spiral arms, global elongation and viewing angles.  On the other hand,
the sum of the effects of spiral arms and global elongation is small,
which means that the two galaxies are close to axisymmetry.  But in
order to quantify this result, more galaxies need to be analyzed in
this way. Work along these lines is in progress (Schoenmakers {\it et
al.}, in prep.).

\section*{acknowledgments}

It is a pleasure to thank Tjeerd van Albada for stimulating
discussions, Konrad Kuijken for a careful reading of an earlier
version of the manuscript, and the referee, Cedric Lacey, for his
insightful comments that significantly improved the paper. F.J.~
Sicking kindly made available the data for both NGC 2403 and NGC 3198.

\vfill
\newpage
\onecolumn
\appendix

\section{Harmonic analysis of the velocity field of a filled gas disk}
\subsection{Deprojection of the line-of-sight velocity}
\label{genap}

Let $R$ and $\theta$ be polar coordinates in the rest frame of the
galaxy and consider a rotating, non-axisymmetric potential in that
frame. It can be written as:
\begin{equation}
\label{potentiaal4}
V(R,\theta,t) = V_0(R) + \sum_m V_m(R)\cos\{m[\theta -\Omega_{p,m}t-
\phi_m(R)]\}.
\end{equation}
\noindent
The potential of a single perturbation in a frame that rotates with the
potential perturbation:
\begin{equation}
  V(R_{\rm orbit},\phi_{\rm orbit}) 
  = V_0(R_{\rm orbit}) + V_m(R_{\rm orbit}) 
    \cos\{m[\phi_{\rm orbit} - \phi_m(R_{\rm orbit})]\},
\end{equation}  
where $V_0(R_{\rm orbit})$ is the unperturbed potential and
$V_m(R_{\rm orbit})\cos\{m[\phi_{\rm orbit} - \phi_m(R_{\rm
orbit})]\}$ is the perturbation.  The subscript "orbit" denotes that
these coordinates describe the point on the closed loop orbit with
guiding centre on $(R_0,\phi_0)$.

The derivation given in this Appendix is also valid for a summation of
perturbations and the final result is a linear addition of the results
for the individual perturbations. There is some coupling between
different terms that occurs when calculating $v_\phi$, equation
(\ref{velos}), but this coupling occurs in second order equations only
and can be ignored in first order calculations.

Following the treatment of Binney \& Tremaine (1987, p.\ 146), we find
that the solution for closed loop orbits is:
\begin{eqnarray}
\label{Req}
R_{\rm orbit} &= R_0 + \delta R =& R_0 \left( 1- \textstyle{\frac{a_{1m}}{2}} \cos\eta -
  a_{2m} \sin\eta \right), \nonumber \\ 
\label {fieq}
\phi_{\rm orbit} &= \phi_0 + \delta \phi =& \phi_0 + \textstyle{\frac{(a_{1m} +
a_{3m})}{2m}} \sin\eta - a_{4m} \cos\eta,
\end{eqnarray}
$R_0$ and $\phi_0$ denote those solutions to the unperturbed problem
that represent a circular orbit ($R=R_0,
\phi=\phi_0=(\Omega_0-\Omega_{p,m})t$). $\Omega_0$ is the circular
frequency: $ \Omega_0 \equiv \sqrt{V_0^{\prime}/R}$ and $\Omega_{p,m}$
is the pattern speed of the particular perturbation.  $\delta R$ and
$\delta \phi$ denote the perturbations on the circular orbit. The
extra terms with respect to the analysis of Binney \& Tremaine arise
from the additional phase term $\phi_m(R_{\rm orbit})$ in the
potential.  Furthermore:
\begin{eqnarray}
\label{S}
\eta = \eta(m,R_{\rm orbit}) &=& m[(\Omega_0-\Omega_{p,m})t -
\phi_m(R_{\rm orbit})] \equiv m[\phi_0 - \phi_m(R_{\rm orbit})], \nonumber \\ 
\label{epsR}
a_{1m} &=& \frac{2}{\Delta_0} \left[ {2\Omega_0 V_m \over
    R_0(\Omega_0-\Omega_{p,m})} + V_m^{\prime}  \right], \nonumber\\ 
\label{epsc}
a_{2m} &=& { m V_m \phi_m^{\prime} (R_{\rm orbit}) \over \Delta_0 }, \\ 
\label{epsv}
a_{3m} &=&\frac{2}{\Delta_0} \left\{ {V_m [(2+m^2)\Omega_0^2 +2(1-m^2)\Omega_0\Omega_{p,m} + m^2
    \Omega_{p,m}^2- \kappa_0^2] \over R_0(\Omega_0-\Omega_{p,m})^2} +
  {(\Omega_0+ \Omega_{p,m} )\over (\Omega_0-\Omega_{p,m})} V_m^{\prime} 
\right\},\nonumber \\ 
\label{epsd}
a_{4m} &=& { 2 \Omega_0 V_m \phi_m^{\prime} (R_{\rm orbit}) \over \Delta_0
  (\Omega_0-\Omega_{p,m})}, \nonumber
\end{eqnarray}
where differentiation with respect to $R$ is denoted as $^{\prime}$,
$\Delta_0 = R_0 \left[ \kappa_0^2 - m^2(\Omega_0-\Omega_{p,m})^2
\right],$ and $\kappa_0^2 \equiv 3\Omega_0^2 + V_0^{\prime \prime}$,
the usual epicycle frequency. These equations break down near
resonances.

The velocities in a frame at rest with respect to the observer are:
\begin{eqnarray}
\label{velos}
  v_R &=& m v_c (1-\omega_m) \left( \textstyle{\frac{1}{2}} a_{1m} \sin\eta -
    a_{2m} \cos\eta \right), \nonumber \\ v_\phi &=& v_c\left\{ 1
    +\textstyle{\frac{1}{2}} [(1-\omega_m)a_{3m} -\omega_m a_{1m}] \cos\eta -[m(1-\omega_m) a_{4m} - a_{2m}]
    \sin\eta \right\} ,
\end{eqnarray}
where $\omega_m \equiv {\Omega_{p,m} \over \Omega_0}$ and $v_c(R_0) = R_0
\Omega_0 $, the circular velocity.

So far we only looked at what happens to a single orbit. A complete
velocity field consists of many orbits and the circular velocity may
be a function of radius. Suppose we are observing the point $(R,\phi)$
in the velocity field. What is the orbit that goes through this point,
in other words, what is the $(R_0,\phi_0)$ that corresponds to
$(R,\phi)$?

Let us write equation (\ref{Req}) as 
\begin{eqnarray}
R_{\rm orbit}(t,R_0,\phi_0) = R_0 \left[ 1 + r(R_0,\phi_0)\right],& \phi_{\rm orbit}(t,R_0,\phi_0) = \phi_0 + p(R_0,\phi_0). 
\end{eqnarray}
Then to first order the coordinates for the guiding centre of the
orbit going through $(R,\phi)$ are
\begin{eqnarray}
\label{guide}
R_0 = \frac{R}{1+r(R,\phi)},&  \phi_0 = \phi - p(R,\phi).
\end{eqnarray}
Then
\begin{eqnarray}
v_c(R_0) &=& v_c(R)(1-r {d\ln v_c(R) \over d\ln R}), \nonumber \\
\omega(R_0) &=& \omega(R)(1-r{d\ln\omega(R)\over d\ln R}). 
\end{eqnarray} 
Inserting these equations in equation \ref{velos} and retaining only
first order terms, we find
\begin{eqnarray}
\label{Rgeneral}
  v_R &=& m v_c(R) [1-\omega_m(R)] \left( \textstyle{{1\over 2}}
  a_{1m} \sin\eta - a_{2m} \cos\eta \right), \nonumber \\
  v_{\phi} &=& v_c(R) \left\{ 1+\textstyle{{1 \over 2}}[ (1-\omega)a_{3m}+(\alpha-\omega_m) a_{1m}]\cos\eta - [m(1-\omega_m)a_{4m} - (1- \alpha) a_{2m}] \sin\eta
 \right\} ,
\end{eqnarray}
where $\alpha= \frac{d\ln v_c(R)}{d\ln R}$. The definition of $\alpha$
is such that the relation between the epicyclic frequency $\kappa_0$
and the circular frequency $\Omega_0$ can be written as $\kappa_0^2
=2(1+\alpha)\Omega_0^2$.

We now project the velocity field on the sky.  The line-of-sight
velocity field is
\begin{equation}
\label{vlos}
v_{\rm los}(R) = [v_R \cos (\theta-\theta_{\rm obs}) - v_{\phi}
\sin(\theta-\theta_{\rm obs})] \sin i = [v_R \cos (\phi-\phi_{\rm
  obs}) - v_{\phi} \sin(\phi-\phi_{\rm obs})] \sin i,
\end{equation}
where $\phi_{\rm obs}$ is the angle between the line $\phi=0$ and the
observer. The angle $i$ is the inclination of the plane of the disk
with respect to the observer. Introduce the variable $ \psi =
\theta-\theta_{\rm obs} + {\pi \over 2}= \phi-\phi_{\rm obs} + {\pi
\over 2}$. This coordinate is zero along the line of nodes.  Then:
\begin{equation}
\label{projector}
  v_{\rm los}(R) = [v_R \sin\psi + v_\phi \cos\psi] \sin i. \nonumber
\end{equation}
Now replace $ \phi \rightarrow \psi + \phi_{\rm obs} - {\pi \over 2} $
in the expressions for $v_R$ and $v_{\phi}$ and expand the
line-of-sight velocity in multiple angles of $\psi$.  Furthermore,
define $ v_* = v_c(R) \sin i$ and assume to first order $\phi_0
\approx \phi$.  After the expansion we find :
\begin{eqnarray}
\label{final}
v_{\rm los} \!\!&=&\!\! c_1 \cos \psi + s_{m-1}\sin(m\!-\!1)\psi +
  c_{m-1}\cos(m\!-\!1)\psi +
  s_{m+1}\sin(m\!+\!1)\psi + c_{m+1}\cos(m\!+\!1)\psi ,
\end{eqnarray}
\noindent
with
\begin{eqnarray}
\label{termsvlosap}
\!\!\!\!\! c_1 \!\!\!\!&=&\!\!\!\! v_* \nonumber \\
s_{m-1} \!\!\!\!&=&\!\!\!\! v_* ( -\textstyle{\frac{1}{4}}\{ [m\!-\!(m\!+\!1)\omega_m+\alpha] a_{1m} +
(1\!-\!\omega_m)a_{3m}\}\sin m\varphi_m \textstyle{+\frac{1}{2}}
\left\{m(1\!-\!\omega_m)a_{4m}+[m(1\!-\!\omega_m)\!-\!1+\alpha]a_{2m}\right\}
\cos m\varphi_m),
\nonumber \\
c_{m-1} \!\!\!\!&=&\!\!\!\! v_*(\textstyle{\frac{1}{4}}\{ [m\!-\!(m\!+\!1)\omega_m +\alpha] a_{1m} +
(1\!-\!\omega_m)a_{3m} \} \cos m\varphi_m \textstyle{+\frac{1}{2}} \left\{m(1\!-\!\omega_m)a_{4m}+[m(1\!-\!\omega_m)\!-\!1+\alpha]a_{2m}\right\} \sin m\varphi_m),\nonumber\\ 
s_{m+1} \!\!\!\!&=&\!\!\!\! v_* (\textstyle{\frac{1}{4}}\{
[m\!-\!(m\!-\!1)\omega_m-\alpha]
a_{1m}\!-\!(1\!-\!\omega_m)a_{3m}\}\sin m\varphi_m \textstyle{+\frac{1}{2}}
\left\{m(1\!-\!\omega_m)a_{4m}-[m(1\!-\!\omega_m)\!+\!1-\alpha]a_{2m}\right\}
\cos m\varphi_m),\\
c_{m+1} \!\!\!\!&=&\!\!\!\! v_* (-\textstyle{\frac{1}{4}}\{ [m\!-\!(m\!-\!1)\omega_m-\alpha] a_{1m}\!-\!(1\!-\!\omega_m)a_{3m}\}
\cos m\varphi_m \textstyle{+\frac{1}{2}} \left\{m(1\!-\!\omega_m)a_{4m}-[m(1\!-\!\omega_m)\!+\!1-\alpha]a_{2m}\right\} \sin m\varphi_m).\nonumber  
\end{eqnarray}
where $ v_* = v_c \sin i$ and $ \varphi_m = \phi_{\rm obs} - {\pi}/{2} -\phi_m(R)$.

\subsection{Deprojection with incorrect viewing angles and incorrect centre}

Now we investigate how the expansion of the line-of-sight velocity
$v_{\rm los} = \sum_{i} c_i(R) \cos i\psi + s_i(R) \sin i\psi$ changes
if an incorrect inclination $\hat i$, position angle $\hat \Gamma$ or
centre $(\hat x,\hat y)$ is chosen. We will derive a general equation
for the line-of-sight velocity which is also valid if a combination of
these incorrect choices occur, since to first order we can simply add
the contributions of the individual incorrect parameters.

Let us start with an incorrect inclination $\hat i = i+ \delta i$,
where $\delta i$ is the deviation from the correct value of the
inclination $i$ and is assumed to be small. Then $\hat q =
\cos(i+\delta i)= q+\delta q$. Using the equations for a projected
circular orbit $x'' = \hat R \cos\hat\psi, y'' = \hat q\hat R\sin
\hat\psi$ and $\cos\psi = x'' / R = x''/ \sqrt{x''^2 + y''^2/ q^2}$ we
find that the relation between $\cos \psi$ and $\cos\hat\psi$ is
\begin{equation}
\label{cospsi}
\cos \psi = (1-\frac{\delta q}{4q})\cos \hat\psi + \frac{\delta q}{4 q} \cos{3 \hat\psi}.
\end{equation}
Or, more general
\begin{equation}
\label{cosgeneral}
\cos a\psi = \cos a\hat\psi + \frac{a \delta q}{4 q} \left(
  \cos(a+2)\hat\psi - \cos(a-2)\hat\psi\right) + {\cal O}(\delta q^2).
\end{equation}
and
\begin{equation}
\label{Rinc}
R = \hat R[1+\frac{\delta q}{2q} (1-\cos 2\hat\psi)]
\end{equation}
Note that the amplitudes of the sine and cosine terms in equation
(\ref{final}) are first order, except the $\cos\psi$ term whose
amplitude is $1$, and the corrections in equation (\ref{cosgeneral})
are first order too. Therefore, only the deprojection of the
$\cos\psi$ term will give rise to first order contributions to the
line-of-sight velocity.

Now define $\hat \Gamma = \Gamma + \delta \Gamma$, where $\Gamma$ is
the correct position angle and $\delta \Gamma$ the deviation from
it. In the same way as for the inclination we find 
\begin{equation}
\label{cosgamma}
\cos\psi = \cos \hat\psi -{\delta\Gamma \over
  4}\left[(3q+\frac{1}{q})\sin\hat\psi+(\frac{1}{q}-q)\sin 3\hat\psi\right].
\end{equation}
More general
\begin{equation}
\label{cosgammagen}
\cos a\psi = \cos a\hat \psi+ \frac{a\delta \Gamma}{2 q} \left\{ \frac{(q^2-1)}{2}
\left[\sin (a-2)\hat\psi + \sin (a+2)\hat\psi \right]-(1+q^2)\sin
a\hat\psi\right\}+{\cal O}(\delta \Gamma ^2).
\end{equation}
and
\begin{equation}
\label{Rgamma}
R = \hat R[1+\textstyle{\frac{1}{2}}(\textstyle{\frac{1}{q}}-q)\sin 2\hat\psi
\delta\Gamma]
\end{equation}
Again, only the deprojection of the $\cos \psi$ term will give rise to
first order effects.

Finally let us consider an incorrect choice of the kinematic
centre. Suppose the centre of the coordinate system is shifted from
$(0,0)$ to $(x',y')$. Let us write this shift as $(\delta x,\delta y)$
where the $x$-axis coincides with the major axis of the projected
orbit. Then
\begin{equation}
\label{coscentre}
\cos\psi = \cos\hat\psi + {\delta x\over 2R} (1-\cos
    2\hat\psi)  - {\delta y\over 2qR} \sin 2\hat\psi.
\end{equation}
In general:
\begin{equation}
\cos a \psi = \cos a\hat\psi + {a \delta x \over {2  R}} [ \cos
  (a-1)\hat\psi - \cos (a+1)\hat\psi ] + {a \delta y\over 2qR} \left[\sin
    (a-1)\hat\psi- \sin (a+1)\hat\psi \right]+{\cal O}(\delta x^2,\delta y^2).
\end{equation}
and 
\begin{equation}
\label{Rcenter}
R = \hat R[1+{\delta x \over \hat R} \cos\hat\psi +{\delta y \over
  q\hat R}\sin\hat\psi]
\end{equation}
Here too only the deprojection of the $\cos\psi$ term is important.

Define $dR = R-\hat R = \hat R[\textstyle{\frac{\delta q}{2q}} (1-\cos
2\hat\psi)+\textstyle{\frac{1}{2}}
(\textstyle{\frac{1}{q}}-q)\delta\Gamma \sin 2\hat\psi +{\delta x
\over\hat R} \cos\hat\psi +{\delta y \over q\hat R}\sin\hat\psi]$ and
substitute the relations between $\psi$ and $\hat\psi$ and $R$ and
$\hat R$ into the expansion of $v_{\rm los}$. To first order $c_i(R) =
c_i(\hat R+dR) = c_i(\hat R) + dR ( d c_i(\hat R)/ dR)$.  Since ${d
c_i / dR}$ is of the same order as $c_i$ and $dR$ is also small (order
$\delta q$, $\delta \Gamma$, $\delta x, \delta y$), the only relevant
term is $d c_1/dR$, the shape of the rotation curve (so this
contribution is zero in the case of a flat rotation curve). Noting
that $\hat R d c_1(\hat R)/dR = v_*(\hat R) \alpha$ we can expand the
line-of-sight velocity in the case of incorrect viewing angles and
centre as:
\begin{eqnarray} 
\label{incorlos}
v_{\rm los} &=& v_*(\hat R) \left\{ (1+\alpha){\delta x\over 2R} - (1-\alpha){\delta q\over 4q}\cos\hat\psi  - {\delta \Gamma \over 4}
\left[(3q+{1\over q})-\alpha ({1\over q}-q)\right] \sin \hat\psi 
  -(1-\alpha){\delta x\over 2R} 
\cos2\hat\psi  \right. \nonumber \\ 
&&\left.
  -(1-\alpha){\delta y\over 2qR} \sin 2\hat\psi+(1-\alpha){\delta q
  \over 4q} \cos 3\hat\psi - (1-\alpha){\delta \Gamma \over 4} ({1\over
  q}-q)\sin 3\hat\psi \right\}+ \sum_{i \ge 0} c_{i}(\hat R) \cos
i\hat\psi + s_i(\hat R) \sin i\hat\psi.
\end{eqnarray}
If this is expanded into new harmonics, $v_{\rm los} = \sum_{i} \hat
c_i(\hat R) \cos i\hat \psi + \hat s_i(\hat R) \sin i \hat\psi$, then $ \hat s_i = s_i, \hat c_i =
c_i $, except for $i \le 3$.

\subsection{Deprojection under the assumption of a circular velocity
  field}

Now we will calculate the errors in $\Gamma,i$ and centre if the
general velocity field is deprojected under the assumption of a
circular velocity field $v_{\rm los}=v_* \cos \hat\psi$. 
As a result, the $\chi^2$ deviation from the best fitting circular
field is
\begin{equation}
\label{chikwad}
\chi^2(\delta q, \delta\Gamma, \delta x, \delta y) \simeq \sum_{i \ge
  0} (\hat c_i^2 + \hat s_i^2) -\hat c_1^2,
\end{equation}
where $\hat s_i$ and $\hat c_i$
are expressed in terms of the $\delta q$, $\delta \Gamma$, $\delta x$ and
$\delta y$ in equation (\ref{incorlos}). The best
fitting values of $\delta \Gamma$ and $\delta q$ are given by
\begin{eqnarray}
\label{bestangles}
\delta q = - \frac{4q c_3}{(c_1+c_3)(1-\alpha)},&& \delta \Gamma = 4q{
    \left[(3q^2+1)-\alpha(1-q^2)\right]s_1 + (1-q^2)(1-\alpha)
    s_3 \over {\left\{ [3q^2+1-(1-q^2)\alpha]^2 + (1-\alpha)^2(1-q^2)^2]\right\} (c_1+c_3)}}.
\end{eqnarray}
Equation (\ref{chikwad}) is minimized with respect to $\delta x$ and
$\delta y$ when 
\begin{eqnarray}
\label{obserm1}
\delta x = 2 R c_2/(1-\alpha), && \delta y = 2qR s_2/(1-\alpha). 
\end{eqnarray}
The resulting expressions for $\hat c_i, \hat s_i$ under a circular
fit are:
\begin{eqnarray}
\label{hatcoef}
\lefteqn{\hat c_0 = c_0 + {1+\alpha \over 1-\alpha}c_2,}&& \nonumber \\
\lefteqn{\hat c_2 = 0,}&& \hskip 4cm \hat s_2 = 0,\nonumber\\
\lefteqn{\hat c_1 = c_1[1\!-\!{\delta q\over 4q}(1\!-\!\alpha)]=c_1+c_3,} && \hskip 4cm \hat s_1 = s_1 - c_1\left[(3q+1/q)-(1/q-q)\alpha\right]\delta \Gamma/4, \\ 
\lefteqn{\hat c_3 = c_3 + (1-\alpha){\delta q\over 4q}c_1=0,} && \hskip 4cm \hat
s_3 = s_3 - c_1(1/q - q)(1-\alpha)
\delta\Gamma /4. \nonumber
\end{eqnarray}
 
\subsubsection{Effect of $m=1$ distortion}
\label{m1}
For an $m=1$ distortion equation (\ref{termsvlosap}) gives the correct
coefficients if the correct centre is chosen:
\begin{eqnarray}
c_0 &=& v_*{\textstyle{1\over 4}} \left[ (1-2\omega_1+\alpha)a_{11}+(1-\omega_1)a_{31}\right]\cos\varphi_1 +{\textstyle{1\over 2}}[(1-\omega_1)a_{41}+(\alpha-\omega_1)a_{21}]\sin\varphi_1, \nonumber \\
c_2 &=& -v_*{\textstyle{1\over 4}} \left[ (1-\alpha)a_{11}-(1-\omega_1)a_{31}\right]\cos\varphi_1 +{\textstyle{1\over 2}}[(1-\omega_1)a_{41}-(2-\omega_1-\alpha)a_{21}]\sin\varphi_1, \\
s_2 &=& v_*{\textstyle{1\over 4}} \left[
  (1-\alpha)a_{11}-(1-\omega_1)a_{31}\right]\sin\varphi_1
+{\textstyle{1\over 2}}[(1-\omega_1)a_{41}-(2-\omega_1-\alpha)a_{21}]\cos \varphi_1. \nonumber
\end{eqnarray}
If the velocity field itself is used to derive the centre, these
equations reduce according to equation (\ref{hatcoef}) to:
\begin{eqnarray}
\hat c_0 &=& v_*{\textstyle{1\over 2(\alpha-1)}}\left\{[(1-\alpha)\omega_1
a_{11}-(1-\omega_1)a_{31}]\cos\varphi_1 - 2 [(1-\omega_1)a_{41}+(\alpha\omega_1-1)a_{21}]\sin \varphi_1\right\} \nonumber,\\
\hat c_2 &=& 0,\\
\hat s_2 &=& 0.\nonumber
\end{eqnarray}
These terms are not affected by an error in $q$ and $\Gamma$.

\subsubsection{Effect of $m=2$ distortion}
\label{m2}
With $m=2$, equation (\ref{termsvlosap}) gives the coefficients if a
correct inclination and position angle are chosen:
\begin{eqnarray}
\label{newcs}
c_1 &=& v_*\left\{1 - \textstyle{\frac{1}{4}} \left[ ( 2-3\omega_2+\alpha) a_{12} +(1-\omega_2)
  a_{32} \right]\cos 2\varphi_2 +
\textstyle{\frac{1}{2}}\left[
  2(1-\omega_2)a_{42}+ (1-2\omega_2+\alpha)a_{22} \right]
\sin 2\varphi_2 \right\}, \nonumber \\ 
s_1 &=& v_*\left\{-\textstyle{\frac{1}{4}} \left[ (
  2-3\omega_2+\alpha) a_{12} +(1-\omega_2) a_{32}\right]
\sin 2\varphi_2 + \textstyle{\frac{1}{2}}\left[
  2(1-\omega_2)a_{42}+ (1-2\omega_2+\alpha)a_{22} \right]
\cos 2\varphi_2 \right\}, \nonumber \\ 
c_3 &=& v_*\left\{-\textstyle{\frac{1}{4}} \left[
  (2-\omega_2-\alpha) a_{12} -(1-\omega_2) a_{32}\right]
\cos 2\varphi_2 + \textstyle{\frac{1}{2}}\left[
  2(1-\omega_2)a_{42}- (3-2\omega_2-\alpha)a_{22} \right]
\sin 2\varphi_2\right\}, \\ 
s_3 &=& v_*\left\{-\textstyle{\frac{1}{4}} \left[ (-2+\omega_2+\alpha)
  a_{12} +(1-\omega_2) a_{32}\right] \sin 2\varphi_2+\textstyle{\frac{1}{2}}\left[
  2(1-\omega_2)a_{42}-(3-2\omega_2-\alpha)a_{22} \right]
\cos 2\varphi_2\right\}.\nonumber
\end{eqnarray}
When the velocity field itself is used to derive the position angle
and inclination, we find using equation (\ref{hatcoef}):
\begin{eqnarray}
\label{longeq}
\hat c_1 &=& v_*\left\{1 + \textstyle{\frac{1}{2}} \left[ (\alpha-\omega_2)a_{12} +
  (1-\omega_2)a_{32}\right] \cos 2\varphi_2 +
  [2(1-\omega_2)a_{42}+(\alpha-1)a_{22}]\sin 2\varphi_2
\right\}, \nonumber \\
\hat s_1 &=&  v_*\left\{-\textstyle{\frac{1}{4}} \left[ (2-3\omega_2+\alpha) a_{12}
    +(1-\omega_2) a_{32}\right]\sin 2\varphi_2+\textstyle{\frac{1}{2}} [2(1-\omega_2)a_{42}+
  (1-2\omega_2+\alpha)a_{22}] \cos 2\varphi_2- \right.\nonumber \\
 && \left. -{ [3q^2\!\!+\!\!1\!\!+\!\!\alpha(q^2\!\!-\!\!1)] (
   \left\{2(1\!\!-\!\!\alpha)(\omega_2\!\!-\!\!\alpha)+2[2\!\!-\!\!\alpha^2\!\!-\!\!4\omega_2\!\!+\!\!\alpha(2\!\!+\!\!\omega_2)]q^2\right\} a_{12}\!\!-\!\![1+q^2\!\!-\!\!\alpha(1\!\!-\!\!q^2)](1\!\!-\!\!\omega_2)
   a_{32}) \over {4[1+2q^2+5q^4+\alpha^2(1-q^2)^2+2\alpha(q^4-1)] }} 
\sin 2\varphi_2 - \right.\nonumber \\
&& \left. -{[3q^2\!\!+\!\!1\!\!+\!\!\alpha(q^2\!\!-\!\!1)] \left\{
  4 [1\!\!+\!\!q^2\!\!-\!\!\alpha(1\!\!-\!\!q^2)](1\!\!-\!\!\omega_2) a_{42} -
  2[1\!\!-\!\!(3\!\!-\!\!4\omega_2)q^2\!\!-\!\!\alpha(1\!\!+\!\!q^2)]
  a_{22}\right\} \over
  {4[1+2q^2+5q^4+\alpha^2(1-q^2)^2+2\alpha(q^4-1)] }}\cos 2\varphi_2\right\}, \nonumber \\ 
\hat c_3 &=& 0, \\
\hat s_3 &=& v_*\left\{-\textstyle{\frac{1}{4}} \left[ (\omega_2+\alpha-2) a_{12}
    +(1-\omega_2) a_{32}\right]\sin 2\varphi_2 +\textstyle{\frac{1}{2}} [2(1-\omega_2)a_{42}-
  (3-2\omega_2-\alpha)a_{22}]\cos 2\varphi_2 +\right.\nonumber\\
&& \left. +{ (1\!\!-\!\!\alpha)(1\!\!-\!\!q^2)(
   \left\{2(1\!\!-\!\!\alpha)(\omega_2\!\!-\!\!\alpha)+2[2\!\!-\!\!\alpha^2\!\!-\!\!4\omega_2\!\!+\!\!\alpha(2\!\!+\!\!\omega_2)]q^2\right\} a_{12}\!\!-\!\![1\!\!+\!\!q^2\!\!-\!\!\alpha(1\!\!-\!\!q^2)](1\!\!-\!\!\omega_2)
   a_{32})   \over 4{[1+2q^2+5q^4+\alpha^2(1-q^2)^2+2\alpha(q^4-1)] }}
  \sin 2\varphi_2 + \right.\nonumber \\
&& \left.+{ (1\!\!-\!\!\alpha)(1\!\!-\!\!q^2) \left\{
  4 [1\!\!+\!\!q^2\!\!-\!\!\alpha(1\!\!-\!\!q^2)](1\!\!-\!\!\omega_2) a_{42} \!\!-\!\!
  2[1\!\!-\!\!(3\!\!-\!\!4\omega_2)q^2\!\!-\!\!\alpha(1\!\!+\!\!q^2)]
  a_{22}\right\}  \over
  {4[1+2q^2+5q^4+\alpha^2(1-q^2)^2+2\alpha(q^4-1)] }} \cos 2\varphi_2\right\}. \nonumber 
\end{eqnarray}
These equations are independent of a small error in the position of
the centre.
For a triaxial non-rotating halo model, in which the potential is scale free, we
have $\omega_2=0$, $\alpha = 0$ (flat rotation curve), the phase of
the perturbation is constant with radius $\varphi_2(R)\sim \mbox{const}$
(therefore $a_{42}=a_{22}=0$),
$a_{32}=2a_{12}=2\epsilon_{\rm pot}$ (FvGdZ) and these
equations simplify to the equations (\ref{fgzterms}). 

\subsubsection{Solid body rotation}

Let us finally consider the effect of solid body rotation, $\alpha =
1$. Equations (\ref{bestangles}) and (\ref{obserm1}) show that in this
case it is impossible to measure the inclination. Indeed, in solid
body rotation, all iso-velocity contours are parallel and each
inclination fits equally well. The position angle on the other hand
can be determined, $\delta \Gamma= {s_1 \over q}$. Centre and systemic
velocity cannot be determined uniquely. Therefore, assumptions for all
these parameters have to be made in order to get a rotation
velocity. It is not possible to interpret harmonic terms under these
conditions in the way we did in this Appendix.
\end{document}